\documentclass[conference]{IEEEtran}
\usepackage{amsmath,amssymb,amsfonts}
\usepackage{algorithmic}
\usepackage{authblk}
\usepackage{textcomp}
\usepackage{xspace}
\usepackage{multirow}
\usepackage{balance}
\usepackage{graphicx}
\usepackage{epstopdf}
\usepackage{multirow}
\usepackage[ruled,vlined]{algorithm2e}
\usepackage[hyphens]{url}
\usepackage{latexsym}
\usepackage{amsthm}
\usepackage{relsize}
\usepackage[nodisplayskipstretch]{setspace}
\usepackage{color}
\usepackage{eucal}
\usepackage{epstopdf}
\usepackage{microtype}
\usepackage{booktabs}
\usepackage[table,xcdraw]{xcolor}
\usepackage{hyperref}
\usepackage[numbers,sort&compress]{natbib}
\usepackage{enumitem}
\usepackage{pifont}
\usepackage{cuted}
\usepackage{flushend}
\usepackage{caption}
\usepackage{subcaption}
\newcommand{\PreserveBackslash}[1]{\let\temp=\\#1\let\\=\temp}
\newcolumntype{M}[1]{>{\PreserveBackslash\centering}m{#1}}
\newcolumntype{C}[1]{>{\PreserveBackslash\centering}p{#1}}
\newcolumntype{R}[1]{>{\PreserveBackslash\raggedleft}p{#1}}
\newcolumntype{L}[1]{>{\PreserveBackslash\raggedright}p{#1}}

\newcommand{\fakepar}[1]{\smallbreak\noindent}

\begin{document}

\makeatletter
\newcommand{\linebreakand}{%
  \end{@IEEEauthorhalign}
  \hfill\mbox{}\par
  \mbox{}\hfill\begin{@IEEEauthorhalign}
}
\newcommand*{\affaddr}[1]{#1} % No op here. Customize it for different styles.
\newcommand*{\affmark}[1][*]{\textsuperscript{#1}}
\newcommand*{\email}[1]{\text{#1}}

\makeatother

\title{Defeating Proactive Jammers Using Deep Reinforcement Learning for Resource-Constrained IoT Networks}

\author{%
Abubakar S. Ali\affmark[1], Shimaa Naser\affmark[1], and Sami Muhaidat\affmark[1,2]\\
\\
\affaddr{\affmark[1]Department of Electrical Engineering and Computer Science,  Khalifa University, Abu Dhabi 127788, UAE}\\
\affaddr{\affmark[2]Department of Systems and Computer Engineering, Carleton University, Ottawa, ON K1S 5B6, Canada}
}

\maketitle

\begin{abstract}
Traditional anti-jamming techniques like spread spectrum, adaptive power/rate control, and cognitive radio, have demonstrated effectiveness in mitigating jamming attacks. However, their robustness against the growing complexity of internet-of-thing (IoT) networks and diverse jamming attacks is still limited. To address these challenges, machine learning (ML)-based techniques have emerged as promising solutions. By offering adaptive and intelligent anti-jamming capabilities, ML-based approaches can effectively adapt to dynamic attack scenarios and overcome the limitations of traditional methods. In this paper, we propose a deep reinforcement learning (DRL)-based approach that utilizes state input from realistic wireless network interface cards. We train five different variants of deep Q-network (DQN) agents to mitigate the effects of jamming with the aim of identifying the most sample-efficient, lightweight, robust, and least complex agent that is tailored for power-constrained devices. The simulation results demonstrate the effectiveness of the proposed DRL-based anti-jamming approach against proactive jammers, regardless of their jamming strategy which eliminates the need for a pattern recognition or jamming strategy detection step. Our findings present a promising solution for securing IoT networks against jamming attacks and  highlights substantial opportunities for continued investigation and advancement within this field.
\end{abstract}


\IEEEpeerreviewmaketitle

%------------------------------------------------------------------%
% Introduction
%------------------------------------------------------------------%
\section{Introduction}
Cognitive radio networks (CRNs) have emerged as a revolutionary paradigm in wireless communication, offering intelligent means to optimizate the available spectrum resources through dynamic channel identification \cite{Haykin2005Cognitive}. Neverthless, the open nature of wireless communication channels exposes CRNs  to potential security breaches, particularly jamming attacks which can degrade  network performance and significantly reduce the throughput  \cite{Fragkiadakis2012Asurvey}. Traditional jamming countermeasures, such as frequency hopping or direct sequence spread spectrum (DSSS), have inherent limitations, especially when confronted with advanced jammers that are capable of detecting and disrupting these techniques \cite{Torrieri2005Principles}. Although, game-theoretical strategies have been explored to address this issue, such techniques assume impractical preconditions like a priori knowledge of the perturbation pattern and can falter when faced with rapidly changing jamming strategies \cite{Xu2018Aone,Noori2020Jamming,Ahmed2017Stackelberg}.

Deep reinforcement learning (DRL), a blend of reinforcement learning and deep learning, has been spotlighted due to its adaptability to dynamic environments and ability to learn from raw data, without the need for pre-existing knowledge. In the context of anti-jamming systems, DRL has been employed in various ways in multiple works. For instance, the authors of \cite{Liu2018Anti} proposed a deep anti-jamming reinforcement learning algorithm (DARLA) that used raw spectrum data as the environmental state, addressing the anti-jamming problem in a dynamic environment. Similarly, the work in \cite{Liu2019Pattern} proposed a sequential deep reinforcement learning algorithm (SDRLA) to improve anti-jamming performance. Other research has introduced wideband autonomous cognitive radios \cite{Machuzak2016Reinforcement}, transformer encoder-like Q-networks \cite{Xu2020Anintelligent}, and unmanned aerial vehicle (UAV) jammers modeled as partially observable Markov decision processes \cite{Gao2019Anti}. Some studies have also used the signal-to-interference-plus-noise ratio (SINR) to enhance anti-jamming techniques \cite{Xiao2018Two, Bi2019Deep}. However, the aforementioned studies relied on supplementary equipment or data such as raw spectrum data or SINR  which can be energy-inefficient and difficult to acquire, rendering them unsuitable for resource-constrained internet-of-things (IoT) networks.\\
\indent In our prior study \cite{Ali2022Deep}, we introduced a novel approach that uses a single vector of clear channel assessment (CCA) information as the state input. This simplifies the environmental state representation, hence, reducing the computational complexity of the neural network. Our previous work also was a departure from the approach presented in \cite{Liu2019Pattern} as it involved a generic DRL agent capable of effectively operating within dynamic jamming pattern environments without requiring a preliminary pattern recognition process. However, despite these capabilites, the CCA-based method faces some challenges, particularly related to the information extraction from WLAN network interface cards (NICs) and its efficacy against random channel hopping jamming. In this paper, we strive to overcome these challenges by proposing an improved anti-jamming scheme. In specific, we exploit a novel radio frequency (RF)-jamming detection testbed \cite{ali2022rf}, utilize the spectrum sensing capabilities of WLAN NICs, and apply ML algorithms to detect and avoid jamming attacks. Additionally, we conduct a comprehensive investigation of different agent alternatives to optimize the anti-jamming performance in dynamic pattern jamming scenarios.

%The paper is organized as follows: The adopted system model and problem formulation are discussed in Sec.\ref{sec:model} and Sec.\ref{sec:problem_formulation}, respectively. The DRL-based anti-jamming strategy is elaborated in Sec.\ref{sec:ch5_DRL-based_approach}. Simulation results and corresponding discussions are provided in Sec.\ref{sec:results}, and the paper concludes with future directions in Sec.~\ref{sec:conc}.

%------------------------------------------------------------------%
% System Model
%------------------------------------------------------------------%
\begin{figure}[ht] 
    \centering
    \includegraphics[width=0.95\columnwidth]{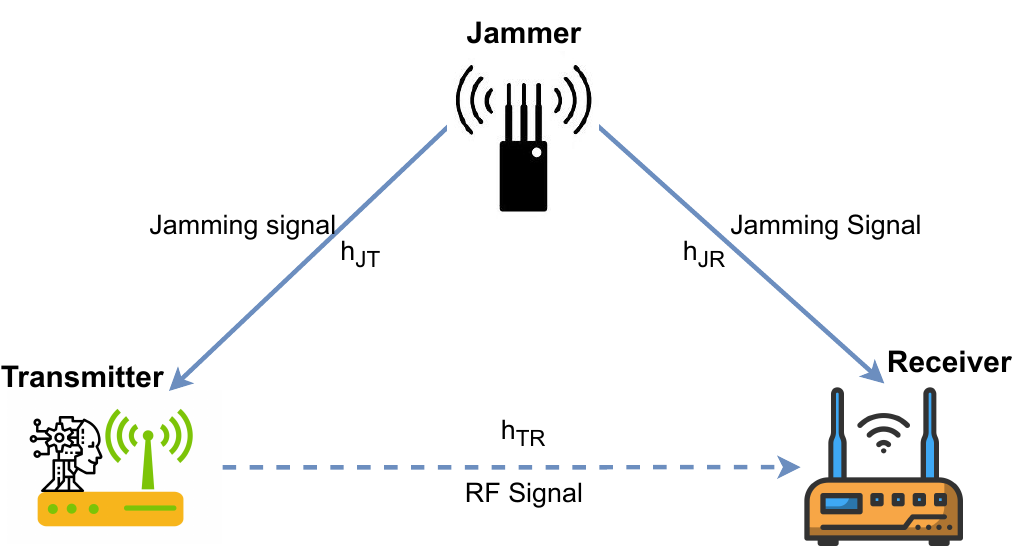}
    \caption{System topology is composed of the transmitter, receiver, and jammer. The transmitter tries to communicate with the receiver in the presence of a jamming attack.}
    \label{fig:sys_model}
\end{figure}
\section{System Model and Formulation}
\label{sec:modelV2}
In this section, we describe the system, jammer, and signal models under jamming attack as illustrated in Fig.~\ref{fig:sys_model}. We consider the UNII-1 band of the 5GHz radio spectrum and assume that the radio environment consists of one user (a transmitter-receiver pair) against one jammer.  A novel aspect of our model is the presence of an agent at the transmission end, which formulates real-time anti-jamming strategies. These strategies are then shared with the receiver through a reliable control link. We also assume that the transmitter possesses broad-band spectrum sensing capabilities \cite{Ali2022Deep}. For ease of analysis, we segment the continuous time into discrete time slots, assuming that both the user and the jammer operate within the same time slot. In each time slot $t$, the user selects a frequency $f_{T,t}$ from the range $\left [ f_L,f_U \right ]$ for data transmission to the receiver, using power $P_{T,t}$. Concurrently, the jammer attempts to interrupt this transmission by selecting a frequency $f_{J,t}$ and power $P_{J,t}$ according to a predefined jamming pattern.

\subsection{Jammer Model}
To investigate proactive jamming attack mitigation, we adopt a range of jamming strategies to effectively counter such threats. Specifically, we employ four distinct approaches: constant, sweeping, random, and dynamic jamming techniques. In this model, we assume that the jammer jams a single frequency $f_{J,t}$ with a varying distance $d_{JT}$ between the jammer and transmitter and varying jamming powers $P_{J,t}$. Given the proactive nature of the jammer, it is assumed to be unaware of the current state of the channel. In the case of the constant jamming strategy, at the beginning of a transmission session, the jammer picks one of the available channels of the RF spectrum to jam consistently. Operating in a manner similar to the constant jammer, the combined jammer possesses the ability to disrupt multiple channels. However, it should be noted that not all channels can be jammed simultaneously by this particular type of jammer. On the other hand,  in the sweeping jamming strategy, the jammer starts jamming the RF spectrum starting from $f_L$ (i.e. $f_{J,t}=f_{L})$ and gradually increasing its jamming frequency until it reaches $f_U$ (i.e. $f_{J,t} = f_U$) in a sweeping fashion. The change of frequency from one to the adjacent occurs at the beginning of each time slot. In contrast, in the random jamming strategy, the jammer randomly selects a frequency $f_{J,t}$ from the set of the available frequencies $\left \{ f_L, \cdots,f_U \right \}$ and jams at the beginning of every time slot. Finally, in the dynamic pattern jamming strategy, the jammer has the capability of selecting one of the three aforementioned jamming strategies (i.e. constant, sweeping, or random) at the beginning of each transmission session. 

%-----------------------------------------------------------
% Transceiver Model
%-----------------------------------------------------------
\subsection{Signal Model}
The received discrete baseband signal $r[n]$  at the receiver after matched filtering and sampling at the symbol intervals can be expressed as follows
\begin{equation}
    r\left [ n \right ]=\sqrt{P^{rx}_T}\;x\left [ n \right ] + \sqrt{P^{rx}_J}\;j\left [ n \right ] + w\left [n  \right ],
\end{equation}
where $x[n]$ and $j[n]$ represent the discrete-time baseband signals transmitted by the transmitter and the jammer, respectively. Furthermore, $w[n]$ denotes the zero-mean additive white Gaussian noise (AWGN) with variance $\sigma ^2$. Finally, $P^{rx}_T$ and  $P^{rx}_J$ represent the average received power from the transmit and the jamming signals, respectively, which can be written as follows
\begin{equation}
    P^{rx}_J = \phi ^{JR} P_{J,t},
\end{equation}
and
\begin{equation}
    P^{rx}_T = \phi ^{TR} P_{T,t},
\end{equation}
where $\phi ^{JR} = \gamma_0 d_{JR}^{-\epsilon}$ and $\phi ^{TR} = \gamma_0 d_{TR}^{-\epsilon}$ are the channel power gains of the jammer-receiver and transmitter-receiver links, respectively. Also, $\gamma_0$ represents the channel power gain at a reference distance of 1m.  $d_{JR}$ and  $d_{TR}$ are the distances of the jammer-receiver and transmitter-receiver links, respectively.  Finally, $\epsilon \geq 2$ denotes the path loss exponent. 

\subsection{Problem Formulation}
The received SINR can be therefore expressed as follows
\begin{equation}
   \Theta = \frac{P^{R}}{P^{rx}_J+\sigma^2},
\end{equation}
where $P^R$ is the power received from the transmitted signal at the receiver. 

Consider $\Theta_{th}$ as the SINR threshold required for successful transmission. The objective at time slot $t$ is to maximize the normalized throughput, defined as $\mathcal{U}(f_{T,t})= \delta (\theta \ge \theta_{th})$, where $\delta(x)$ is a function that equals 1 if $x$ is true, and 0 otherwise. 
\section{Proposed DRL-Based Approach}
\label{sec:ch5_DRL-based_approachV2}
In this section, we introduce a DRL-based anti-jamming scheme that obtains its state information by scanning the entire spectrum. 
\subsection{MDP Formulation}
We utilize the received power feature from the generated dataset to represent the state vector $\mathbf{P_t}$. Specifically, the state vector is represented as $\mathbf{P_t} = [p_{t,1}, p_{t,2}, \cdots, p_{t,N_c}]$, where $p_{t,i}$ is the received power at time $t$ for frequency $i$. The size of the state space is $\left |\mathcal{S} \right | = N_c$. In our formulation, the action $a_t \in \{f_1,f_2,\cdots, f_{N_c}\}$ represents the selection of frequency $i$ at time slot $t$. Similarly, the action space size is $\left |\mathcal{A} \right | = N_c$. The transmitter-receiver pair aims to achieve successful transmission with a low channel switching cost $\Gamma$. Therefore, the reward at time slot $t$ can be expressed as

\begin{equation}\label{eq:ch_reward}
r_t = \begin{cases}
\mathcal{U}(f_{T,t}) -\Gamma \delta (a_t\neq a_{t-1}) & \text{ if } f_{T,t}\neq f_{J,t} \\
0 & \text{ if } f_{T,t}= f_{J,t}.
\end{cases}
\end{equation}

The reward function presented in (\ref{eq:ch_reward}) takes into account the throughput factor and ignores the energy consumption factor. This is due to the fact that in the current anti-jamming strategy, the transmit power is fixed. Furthermore, the normalization of the reward values to 1 and 0 is valid since the considered jammer is proactive. Based on this, upon obtaining the reward $r_t$, the environment transitions to the next state $s_{t+1}$ based on a transition probability  $p(s_{t+1}|s_t, a_t)$. This probability represents the likelihood of transitioning from state $s_t$ to state $s_{t+1}$ given the action $a_t$. The initial state is denoted by $s_0$ and the terminal state is the state at which the agent ceases decision-making, which is denoted by $s_T$. The goal of the agent is to find the optimal policy, $\pi{(s)} = \arg\max_{a} Q(s, a)$, that maps the state to the best action. The optimal policy is found by learning the optimal action-value function, $Q^*(s, a)$, using an RL algorithm such as DRL.

\begin{figure}
    \centering
    \includegraphics[width=1\linewidth]{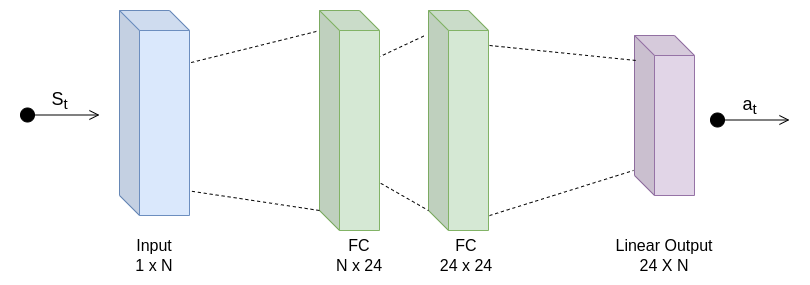}
    \caption{Architecture of the proposed DDQN Q-network.}
    \label{fig:QnetFig}
\end{figure}
\begin{figure}
    \centering
    \includegraphics[width=\columnwidth]{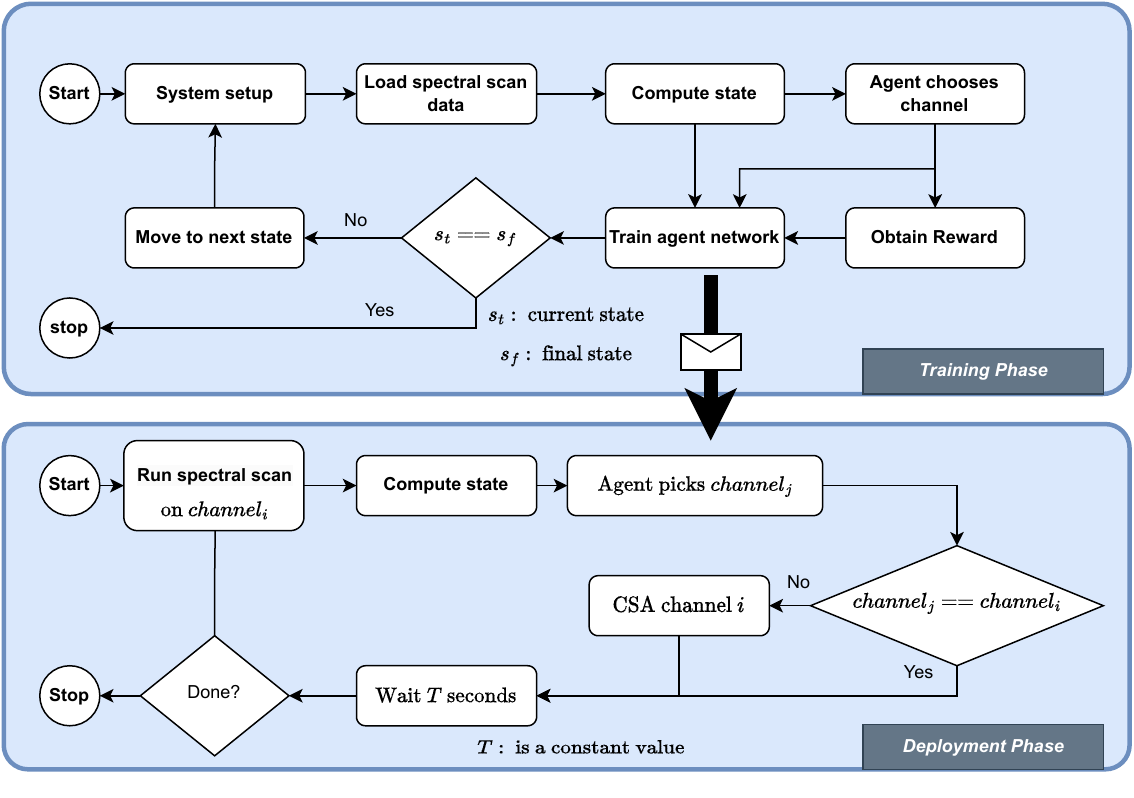}
    \caption{Overview of the training and deployment phases of the proposed DRL-based anti-jamming approach.}
    \label{fig:ch5_DRLApproach}
\end{figure}

\subsection{Agent Design}
We train five different agents to determine the most suitable strategy for power-constrained devices. These agents include DQN, DQN with fixed targets, DDQN, Dueling DQN, and DDQN with prioritized replay. Each agent has a unique combination of neural network architecture, experience replay mechanism, and target network update frequency. By training and evaluating the performance of these agents, we aim to identify the most appropriate approach for power-constrained devices in effectively countering proactive jamming attacks.

\subsubsection{DQN}
The DQN algorithm is a model-free, online, off-policy RL method in which a value-based RL agent is employed to train a Q-network that estimates and returns future rewards \cite{Mnih2015Human, Mnih2013Playing}. The selection of this type of agent is motivated by the fact that our observation space is continuous, and our action space is discrete. Our DQN algorithm implementation is presented in Algorithm \ref{alg:ch5_DQN}.

The implemented DQN agent uses a function approximator in the form of a neural network, whose weights $\theta_{Q}$ are updated with every iteration. The Q-network is used to determine the Q-value of the action. The Q-network comprises two hidden layers, as illustrated in Fig. \ref{fig:QnetFig}, and a ReLU activation function $f(x)=\textup{max}(0,x)$ is chosen \cite{Nair2010Rectified}. The experience reply buffer $\mathcal{D}$ stores the agent's experience, which is the transition pair at time-step $t$ and is defined as $(s_{t},a_{t},r_{t},s_{t+1})$.

The stochastic gradient descent (SGD) algorithm \cite{Bottou2008The} is used during training to update the weights $\theta_{t}$ at every time-step $t$.

\begin{algorithm}
\SetAlgoLined
$\bold{Initialize}$
 $\theta_{Q}, \epsilon_{t} =1, \delta, i=j=0, \textup{ and } K$\;
 \While{j $< \left |\mathcal{E}  \right |$}{
  set $s_{t} = s_{t_{0}}$\;
    \While{t $< \left |\mathcal{T}  \right |$}{
        $X_t \sim U\left ( 0,1 \right )$\;
        \eIf{$X_{t} < \epsilon_{t}$}{
            $a_{t} = \textup{random}(1,\cdots, N_c)$\;
            }{
            $a_{t} = \underset{a_{t}}{\textup{arg max}}Q(s_{t},a{t}\mid \theta_{Q})$\;
            }
        $a_{t} \mapsto \bold{T}$\;
        Obtain $r_{t}$ and $s_{t+1}$\;
        Store the experience $[s_{t}, a_{t},r_{t},s_{t+1}]$ in $\mathcal{D}$\;
        Sample a random mini-batch of $K$ experiences from $\mathcal{D}$\;
        \eIf{$s_{t} == s_{t_{f}}$}{
            $y_{t} = r_{t}$\;
        }{
            $y_{t}=\mathbb{E}_{s_{t},a_{t}}[r_{s_{t},s_{t+1},a_{t}}+\gamma Q_{\pi}(s_{t+1},a_{t_{\textup{max}}}\mid \theta_{Q})\mid s_{t},a_{t}]$\;
        }
        Update Q-network parameters $\theta_{Q}=\theta_{t}-\eta \nabla L_{t}(\theta_{t})$\;
        where $L_{t}(\theta_{t})=\mathbb{E}_{s_{t},a_{t}}[(y_{t}-Q_{\pi}(s_{t},a_{t};\theta_{t}))^{2}]$\;
        Update the exploration rate $\epsilon_{t+1}=\epsilon_{t}-\delta$\;
        Set $s_{t}=s_{t+1}$\;
        $t=t+1$\;
    }
    $j=j+1$\;
 }
 $\mathbf{Output}$ optimal policy $\overset{*}{\pi}$\;
 \caption{DQN Algorithm for Anti-Jamming.}
 \label{alg:ch5_DQN}
\end{algorithm}
    
\subsubsection{DQN with Fixed Targets} 
This variant of DQN updates the target network less frequently, reducing the risk of oscillations and instability during learning. The algorithm is similar to the DQN, but the target network is updated less frequently. This can be achieved by increasing the value of $C$ (the number of steps between target network updates). The neural network architecture and other components remain unchanged from the DQN architecture.

\subsubsection{DDQN} 
The Double Deep Q-Network (DDQN) is an improvement over DQN that reduces the overestimation of Q-values by using two separate networks to estimate the current and target Q-values. The neural network architecture and other components remain unchanged from the DQN architecture.

\subsubsection{Dueling DQN}
This algorithm is similar to the DQN, but with a different neural network architecture that decouples the estimation of state values and action advantages, potentially leading to better performance and stability. To implement this, the architecture of the Q-network in Fig. \ref{fig:QnetFig} is modified to include two separate streams for state values and action advantages, and then these streams are combined to obtain the final Q-values. The other components remain unchanged from the DQN architecture.

\subsubsection{DQN with Prioritized Replay} 
This approach combines DQN with prioritized experience replay, which samples more important experiences more frequently during learning, potentially improving learning efficiency. To implement this, the uniform sampling of experiences from the replay buffer $\mathcal{D}$ is replaced with prioritized sampling based on the absolute TD-error of each experience. Additionally, the loss function $L_{t}(\theta_{t})$ is updated to include importance-sampling weights to correct for the bias introduced by the prioritized sampling. The neural network architecture remains unchanged from the DQN architecture.

\subsection{Training and Deployment of the Agent}
\label{sec:ch5_DRL-training}
In this section, we detail the training and deployment of our proposed DRL-based anti-jamming approach, which aims to mitigate jamming attacks in power-constrained devices. Fig. \ref{fig:ch5_DRLApproach} presents an overview of the training and deployment phases of the proposed DRL-based anti-jamming approach. The training phase involves the setup of the system, loading the corresponding data from the spectral scan dataset, obtaining the \textit{received power (dBm)} feature of each channel, and training the agents based on the reward value obtained from the selected channel. At the beginning, a system setup is made to specify the type of jammer (i.e., sweeping, random, constant, or dynamic pattern jammers), the jamming power, and the distance. Based on this setup, the corresponding data is loaded from the spectral scan dataset. Depending on the type of jammer, the \textit{received power (dBm)} feature of each channel is obtained. For instance, if the jammer is constant, and the jamming frequency is 5180 MHz at 20 cm with a jamming power of 10 dBm, then the dataset with the corresponding \textbf{filename} will be loaded. This ensures that the 5180 MHz frequency will have the highest received power compared to the other frequencies. Based on this state information, the agent will select a channel and receive a reward value based on the selected channel, as defined in (\ref{eq:ch_reward}). Using this reward value, the agent's network is trained and then the environment transitions to the next state. It is worth noting that  this process repeats until convergence or a terminal state is reached.

During the deployment phase, the trained agent is implemented within the environment it was originally trained on. However, in this phase, the agent does not undergo further training as it exploits the knowledge gained from the training phase. Given a system setup and the current channel $f_{T,t}$, the agent takes in the state vector, which describes the whole spectrum, as input and selects the best channel $f_{T,t+1}$ to switch to. If the selected channel $f_{T,t+1}$ is the same as the current channel $f_{T,t}$, then transmission continues on $f_{T,t}$. If $f_{T,t+1} \neq f_{T,t}$, a channel switch announcement (CSA) is carried out, and the subsequent transmission switches to $f_{T,t+1}$. This process keeps repeating until all data is transmitted or the terminal state is reached.

%------------------------------------------------------------------%
% Numerical and Simulation Evaluations
%------------------------------------------------------------------%
\section{Results and Discussions}
\label{sec:results}
To evaluate the proposed DRL-based anti-jamming solution, we aim to investigate its performance under dynamic pattern jamming, where the jammer randomly selects one of the three jamming patterns namely, sweep, random, and combined at the beginning of each transmission session. This evaluation is important as our primary objective is to develop a generic anti-jamming agent capable of mitigating various jamming patterns. We perform the simulations using a custom-based simulator designed based on the collected dataset in \cite{ali2022rf}. Also, unless otherwise stated, the simulation parameters used in our study are presented in Table~\ref{tab:proposed_simulation_params}. Furthermore, we tune the hyper-parameters of the proposed DRL-based anti-jamming scheme during training to achieve a good policy for the agent, as shown in Table~\ref{tab:RL_params}. Finally, we investigate the effects of the $\Gamma$ parameter on the total throughput of the proposed framework, and we compare the results obtained by using different values of $\Gamma$.

\begin{table}[ht]
    \caption{Simulation Parameters}
    \renewcommand{\arraystretch}{1.1}
    \centering
    \footnotesize
    \begin{tabular}{ccc}
    \toprule
        \bfseries Parameter & \bfseries Value \\
        \midrule
        RF spectrum band & 5GHz UNII-1 \\
        Bandwidth of communication signal & 20 MHz\\
        Bandwidth of jamming signal & 20 MHz\\
        Number of channels $N_c$ & 8\\
        Initial channel center frequency $f_{T,0}$ & 5.180 GHz\\
        Distance between channel frequencies & 20 MHz \\
        Distance between jammer and transmitter $d_{JT}$ & 20 cm\\
        Jamming power $P_{J,t}$ & 10d Bm \\
    \bottomrule
    \end{tabular}
    \label{tab:proposed_simulation_params}\vspace{-0.2cm}
\end{table} 

\begin{table}[htbp]
    \caption{DRL Hyper-parameters.}
    \footnotesize
    \centering
    \begin{tabular}{ccc}
    % \toprule
    \hline
    \textbf{Parameter}	& \textbf{Value} \\
    % \midrule
    \hline
    Number of training episodes $\left |\mathcal{E}  \right |$ & 100\\
    Number of testing episodes $\left |\mathcal{E}  \right |$ & 100\\
    Number of time-steps $\left |\mathcal{T}  \right |$ & 100\\
    Discount factor $\gamma$ & 0.95\\
    Initial exploration rate $\zeta$ & 1\\
    Exploration decay $\delta$ & 0.005\\
    Minimum exploration rate $\zeta_{\textup{min}}$ & 0.01\\
    Experience buffer size $\mathcal{D}$ & 10000\\
    Minimum batch size $K$ & 32\\
    Averaging window size & 10\\
    Early termination criterion & Average reward = 90\\
    Channel switching cost $\Gamma$ & [0, 0.05, 0.1, 0.15]\\
    \hline
    \end{tabular}
    \label{tab:RL_params}
\end{table}

\begin{figure}
  \begin{subfigure}[t]{.49\columnwidth}
    \centering
    \includegraphics[width=\linewidth]{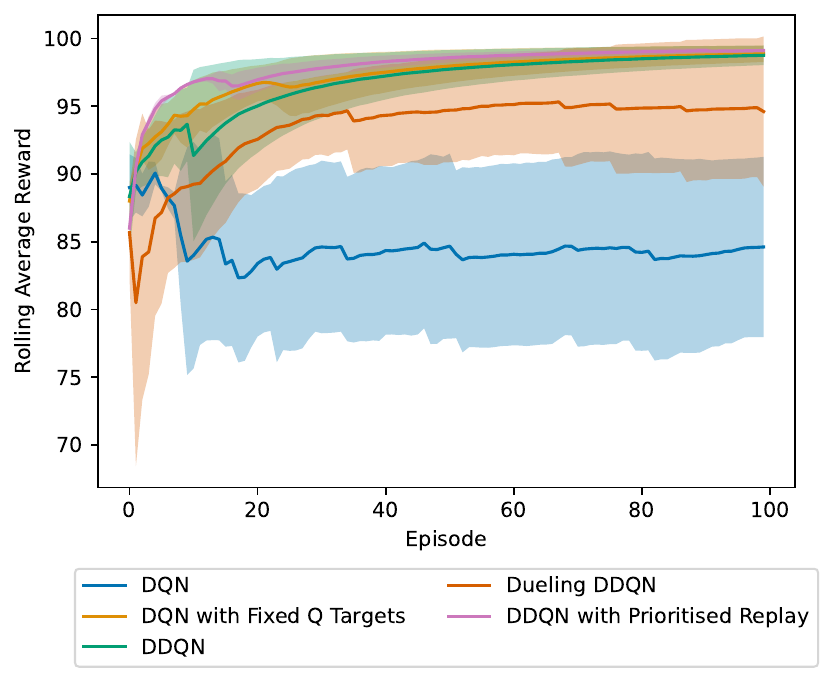}
    \caption{$\Gamma = 0.00$.}
  \end{subfigure}
  \hfill
  \begin{subfigure}[t]{.49\columnwidth}
    \centering
    \includegraphics[width=\linewidth]{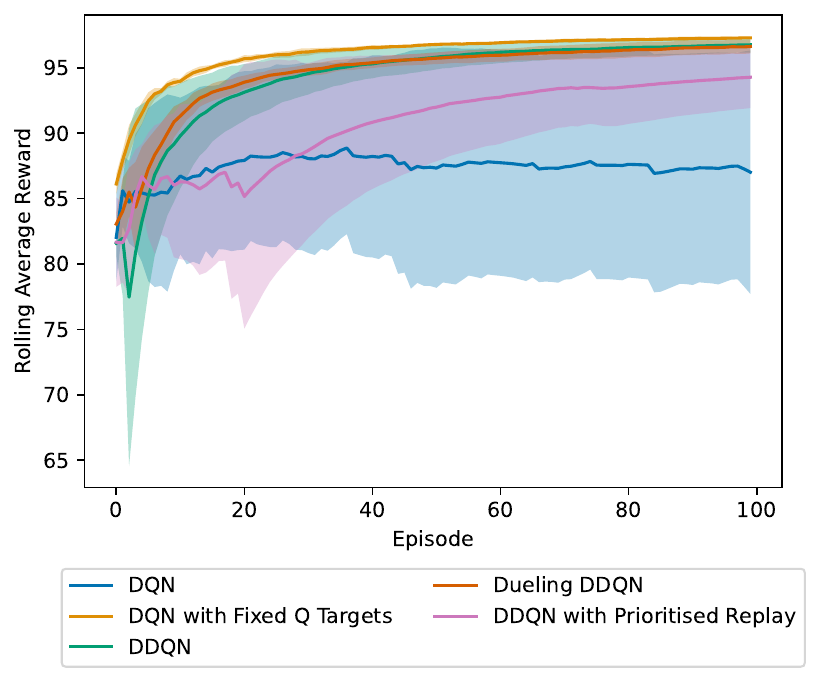}
    \caption{$\Gamma = 0.05$.}
  \end{subfigure}

  \medskip

  \begin{subfigure}[t]{.49\columnwidth}
    \centering
    \includegraphics[width=\linewidth]{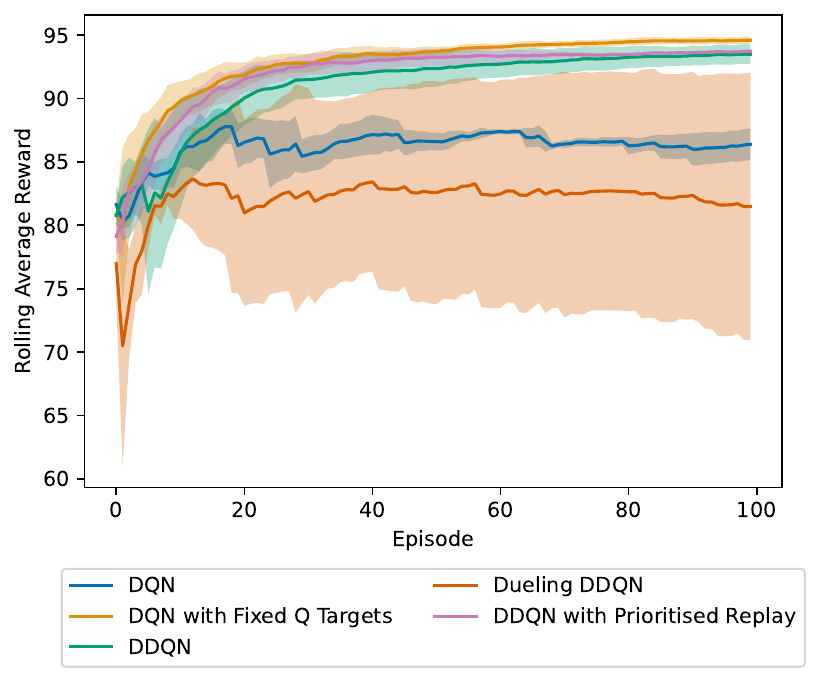}
    \caption{$\Gamma = 0.10$.}
  \end{subfigure}
  \hfill
  \begin{subfigure}[t]{.49\columnwidth}
    \centering
    \includegraphics[width=\linewidth]{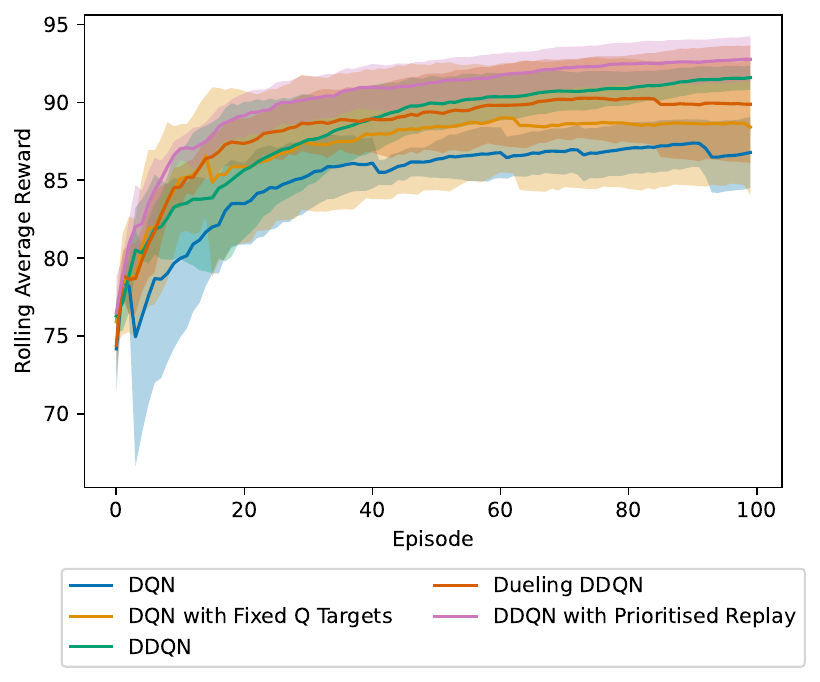}
    \caption{$\Gamma = 0.15$.}
  \end{subfigure}
  \caption{Learning performance of the investigated DRL-based anti-jamming agents under dynamic pattern jamming with $\Gamma = 0,0.05,0.1,0.15.$}
  \label{fig:ch5_learningProposed}
\end{figure}

Fig.~\ref{fig:ch5_learningProposed} depicts the learning performance of the DRL-based anti-jamming agents under dynamic pattern jamming, with different values of $\Gamma$. We observe that DQN with fixed Q-targets, DDQN, and DDQN with prioritized replay achieve a mean reward of approximately 100, while Dueling DQN achieves a mean reward of around 95. However, the DQN agent only manages to obtain a mean reward of approximately 86, and this failure persists for all values of $\Gamma$. Unlike in our prior work, \cite{Ali2022Deep}, in this work all the agents were able to learn the dynamics of the system and evade the jammer. Importantly, we note that all the trained DRL agents, except for DQN, can learn a policy to escape the dynamic pattern jamming. Moreover, we observe that for all types of jammers, the DRL agents can make intelligent channel selection decisions to evade jamming. Interestingly, the DDQN with prioritized replay achieved the most stable learning convergence across all values of $\Gamma$. 

In Fig.~\ref{fig:ch5_throughputProposed}, we present the normalized mean throughput of the legitimate user under various jamming patterns. We observe that, for all values of $\Gamma$, all the evaluated agents, except DQN, have the ability to completely evade dynamic pattern jamming. Moreover, for all agents, we observe a reduction in throughput as the value of $\Gamma$ increases, with a greater reduction for higher values of $\Gamma$. As seen in the case of the learning performance, the DDQN with prioritized replay achieved a consistently high throughput over all values of $\Gamma$.

The impact of $\Gamma$ on the channel switching behavior of the agents is demonstrated in Fig.~\ref{fig:ch5_cscProposed}. It is observed that the agents switch channels 100\% of the time, regardless of the values of $\Gamma$. This indicates that in order to evade dynamic pattern jamming, the agents develop a policy that maps the states to the optimal action and ignores the jamming pattern. This leads to continuous channel switching even under values of $\Gamma > 0$. In other words, the agents choose to be penalized by the channel switching cost and experience a reduction in overall throughput instead of remaining on a single channel and losing 1/8 of their total throughput.

\begin{figure}
    \centering
    \includegraphics[width=\columnwidth]{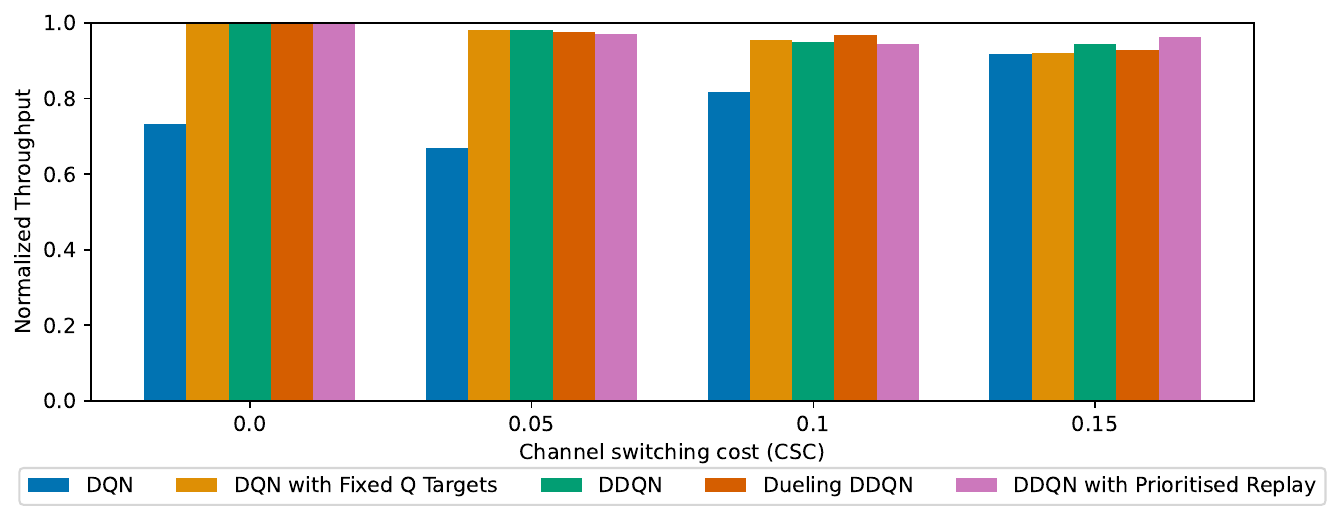}
    \caption{Normalized throughput performance of the DRL-based anti-jamming agent under dynamic pattern jamming.}
    \label{fig:ch5_throughputProposed}
\end{figure}

\begin{figure}
    \centering
    \includegraphics[width=\columnwidth]{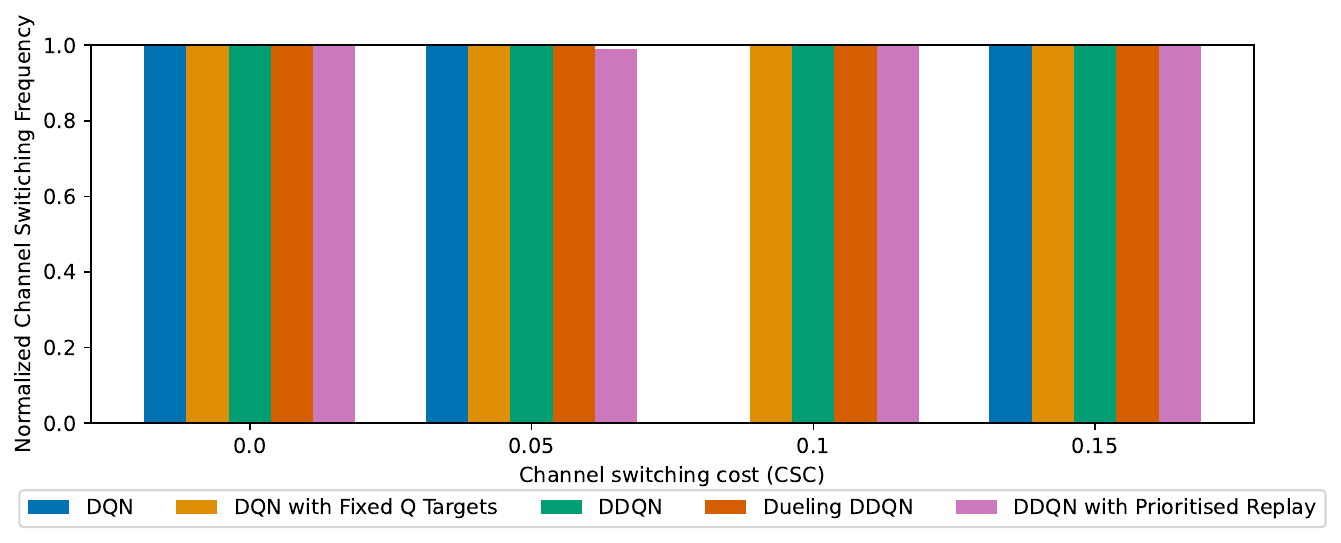}
    \caption{Impact of channel switching cost ($\Gamma$) on the DRL-based anti-jamming agent under dynamic pattern jamming.}
    \label{fig:ch5_cscProposed}
\end{figure}

Finally, we study the convergence times and inference speeds of the five DRL agents as shown in Table \ref{tab:comparison_times}. During training, the DQN agent demonstrated the fastest convergence speed among all the agents, with an average convergence time of 388.28 seconds. The speed of convergence and inference in DRL agents is determined by the complexity of the learning algorithm and the efficiency of the exploration strategy. DQN, with its simpler learning algorithm and efficient exploration, converges faster. On the other hand, DDQN with prioritized replay memory involves more complex computations and a more sophisticated memory management system, which slows down both the convergence and the inference speed.

\begin{table}[h!]
  \centering
  \footnotesize
  \caption{Comparison of the Convergence and inference times for the five Agents. The results are present in the format of \textit{mean} ($\pm$ \textit{std.}) obtained from 10-folds.}
  \label{tab:comparison_times}
  \resizebox{\columnwidth}{!}{
  \begin{tabular}{ccc}
    \toprule
    \textbf{Agent} & \textbf{Convergence Time (sec)} & \textbf{Inference Speed (KHz)} \\
    \midrule
    \textbf{DQN} & 388.28 ($\pm$ 3.62) & 507.23 ($\pm$ 4.30) \\
    \textbf{DQN with Fixed Targets} & 405.37 ($\pm$ 1.74) & 472.43 ($\pm$ 2.15) \\
    \textbf{DDQN} & 457.42 ($\pm$ 3.26) & 437.78 ($\pm$ 3.43) \\
    \textbf{Dueling DQN} & 405.79 ($\pm$ 6.54) & 464.58 ($\pm$ 2.87) \\
    \textbf{DDQN with Prioritized Replay} & 532.85 ($\pm$ 3.91) & 382.31 ($\pm$ 5.25) \\
    \bottomrule
  \end{tabular}}
\end{table}

Overall, all the algorithms investigated showed good performance in jamming detection and avoidance. The inference speed of the algorithms varied, with DQN being the fastest during training. Among all DRL-based approaches, DDQN with prioritized replay memory offers the best trade-off between throughput and speed.
%------------------------------------------------------------------%
% Conclusions
%------------------------------------------------------------------%
\section{Conclusions}
\label{sec:conc}
This paper investigates the intelligent anti-jamming problem within a dynamic jamming environment. In our endeavor to construct a more practical scheme, we incorporated a jamming detection testbed and jamming data acquired from actual WLAN network interface cards. Utilizing this dataset, we developed a custom simulation and introduced a DRL agent with a fully connected neural network architecture to navigate the intricate decision-making problem inherent to anti-jamming. With our proposed scheme, the agent is capable of learning the most effective anti-jamming strategy through a continuous process of trial and error, testing various actions, and observing their environmental impact. We used simulation results from a variety of environmental settings to corroborate the effectiveness of the proposed DRL-based anti-jamming scheme. It's important to note, however, that a high-power wideband jammer leaves no room for evasion. Consequently, future research will involve creating an anti-jamming technique focused on confronting the jammer at the same frequency, as opposed to evasion or concealment.

% References section
%\bibliographystyle{IEEEtran}
%\bibliography{references.bib}
% Generated by IEEEtran.bst, version: 1.14 (2015/08/26)

% that's all folks
\end{document}